\documentclass[preprint,preprintnumbers,showpacs,aps,prd,amssymb]{revtex4-1}

\usepackage{graphicx}
\usepackage{bm}
\usepackage{amsmath}

\def\calA{{\cal A}}
\def\calH{{\cal H}}
\def\calL{{\cal L}}

\def\calO{{\cal O}}
\def\calS{{\cal S}}
\def\calU{{\cal U}}
\def\dU{{d_\calU}}

\def\nn{\nonumber}
\def\SM{{\rm SM}}
\def\Br{{\rm Br}}

\begin{document}
\title{Constraints on Unparticles from $B_s\to\mu^+\mu^-$}
\author{Jong-Phil Lee}
\email{jongphil7@gmail.com}
\affiliation{Division of Quantum Phases $\&$ Devices, School of Physics, Konkuk University, Seoul 143-701, Korea}

\begin{abstract}
Unparticle contributions to the recently measured decay mode $B_s\to\mu^+\mu^-$ are analyzed.
We consider only the scalar unparticles because vector unparticles are expected to provide negligible contributions.
Assuming that the relevant coupling constants are real, we present allowed regions of coupling constants 
and the scaling dimension of the scalar unparticle.
While the measured value of the branching ratio is very close to the standard model predictions, 
one cannot exclude the possible contributions from unparticles.
\end{abstract}

\maketitle
Recently the LHCb collaboration reported the first evidence for the decay $B_s\to\mu^+\mu^-$ and an upper limit on
$B_d\to\mu^+\mu^-$ as \cite{LHCb}
\begin{eqnarray}
 \Br(B_s\to\mu^+\mu^-)&=&\left(3.2^{+1.5}_{-1.2}\right)\times 10^{-9}~,\label{EXP}\\
 \Br(B_d\to\mu^+\mu^-)&<&9.4\times 10^{-10}~.\label{BdEXP}
\end{eqnarray}
The result is quite consistent with the standard model (SM) predictions\cite{SM}
\begin{eqnarray}
 \Br(B_s\to\mu^+\mu^-)_\SM&=&(3.23\pm 0.27)\times 10^{-9}~,\label{SMBs}\\
 \Br(B_d\to\mu^+\mu^-)_\SM&=&(1.07\pm0.10)\times 10^{-10}~.\label{SMBd}
\end{eqnarray}
The decay of $B_s\to\mu^+\mu^-$ is very sensitive to new physics because in the SM the process occurs only through
the loop contributions.
However, it would be too early to declare that there are no new physics at all.
Implications of the new observation in view of new physics can be found in \cite{Arbey,Guadagnoli,Buras}.
In this paper, we examine the unparticle effects on $B_s\to\mu^+\mu^-$ decay.
\par
Unparticle is a hypothetical concept associated with the scale invariance at high energy scales \cite{Georgi}.
According to the unparticle scenario there is a scale-invariant hidden sector,
and it couples to the SM particles very weakly at high energy scale $\Lambda_\calU$.
When seen at low energy, the hidden sector behaves in different ways from ordinary particles. 
That's the reason why the stuff is called as {\em unparticles}.
In other words, unparticles behave like a fractional number of particles.
\par
Consider a ultraviolet (UV) theory in the hidden sector at some high energy $\sim M_\calU$
with the infrared (IR)-stable fixed point.
It is quite convenient to describe the interaction between the UV theory and the SM sector in an
effective theory formalism.
Below the scale of $M_\calU$, a UV operator $\calO_{\rm UV}$ interacts with an SM operator $\calO_{\rm SM}$ through 
$\calO_{\rm SM}\calO_{\rm UV}/M_\calU^{d_{\rm SM}+d_{\rm UV}-4}$,
where $d_{\rm UV(SM)}$ is the scaling dimension of $\calO_{\rm UV(SM)}$.
Through the renormalization flow, one can go down to a new scale $\Lambda_\calU$.
It appears through the dimensional transmutation where the scale invariance emerges.
Below $\Lambda_\calU$ the theory is matched onto the above interaction with
the new unparticle operator $\calO_\calU$ as
\begin{equation}
C_\calU\frac{\Lambda_\calU^{d_{\rm UV}-\dU}}{M_\calU^{d_{SM}+d_{\rm UV}-4}}
\calO_{SM}\calO_{\calU}~,
\end{equation}
where $\dU$ is the scaling dimension of $\calO_\calU$ and $C_\calU$ is the
matching coefficient.
The value of $\dU$ is not constrained to be integers because of the scale invariance.
This unusual behavior of unparticles is reflected on the phase space of $\calO_\calU$.
The spectral function of the unparticle is given by
the two-point function of $\calO_\calU$ as
\begin{eqnarray}
\rho_\calU(P^2)&=&\int d^4x~e^{iP\cdot x}
\langle 0|\calO_\calU(x)\calO_\calU^\dagger(0)|0\rangle\nn\\
&=&
A_{\dU}\theta(P^0)\theta(P^2)(P^2)^{\dU-2}~,
\label{rhoU}
\end{eqnarray}
where
\begin{equation}
A_{\dU}=\frac{16\pi^2\sqrt{\pi}}{(2\pi)^{2\dU}}
\frac{\Gamma(\dU+\frac{1}{2})}{\Gamma(\dU-1)\Gamma(2\dU)}~,
\end{equation}
is the normalization factor.
The corresponding phase space is
\begin{equation}
d\Phi_\calU(P)=\rho_\calU(P^2)\frac{d^4P}{(2\pi)^4}
=A_{\dU}\theta(P^0)\theta(P^2)(P^2)^{\dU-2}\frac{d^4P}{(2\pi)^4}~,
\label{dPhi}
\end{equation}
and the propagator is given by
\begin{equation}
 \int d^4x~e^{iP\cdot x}
\langle 0|T\calO_\calU(x)\calO_\calU^\dagger(0)|0\rangle
=\frac{iA_\dU}{2\sin\dU\pi}\frac{e^{-i\phi_\dU}}{(P^2+i\epsilon)^{2-\dU}}~,
\label{propa}
\end{equation}
where $\phi_\dU=(\dU-2)\pi$.
\par
$B$ physics is a good test bed for the unparticle effects \cite{Geng,Mohanta2},
including $B_s$-$\bar{B}_s$ mixing \cite{Mohanta1,Lenz,Parry,jplee}
(see also \cite{Li,Chen} for meson mixing).
One reason is that unparticles can contribute to the flavor changing neutral current at tree level.
For decays of $B_s\to\ell^+\ell^-$, the scalar unparticle can contribute generally through
\begin{equation}
 \calL=\sum_i\left[
  C_q^i\calO_q^i\calO_\calU+D_q^i(\calO_q^i)_\mu\partial^\mu\calO_\calU
+C^i_\ell\calO_\ell\calO_\calU+D_\ell^i(\calO_\ell^i)_\mu\partial^\mu\calO_\calU\right]~,
\end{equation}
where $C_{q,\ell}^i$ and $D_{q,\ell}^i$ are coefficients. 
The quark operators are
$\calO_q^i={\bar q}q,~ {\bar q}\gamma_5 q$, and
$(\calO_q^i)_\mu={\bar q}\gamma_\mu q,~ {\bar q}\gamma_\mu\gamma_5 q$,
while the leptonic operators are
$\calO_\ell^i={\bar\ell}\ell,~{\bar\ell}\gamma_5\ell$ and
$(\calO_\ell^i)_\mu={\bar\ell}\gamma_\mu\ell,~{\bar\ell}\gamma_\mu\gamma_5\ell$.
\par
For simplicity we only consider the left-handed currents coupled to scalar unparticles by the following Lagrangian
\begin{equation}
 \calL_\calU
=\frac{c_q}{\Lambda_\calU^\dU}\bar q'\gamma_\mu(1-\gamma_5)q~\partial^\mu\calO_\calU
  +\frac{c_\ell}{\Lambda_\calU^\dU}\bar\ell'\gamma_\mu(1-\gamma_5)\ell~\partial^\mu\calO_\calU~,
\label{LU}
\end{equation}
where $c_{q,\ell}$ are dimensionless couplings.
We assume that $c_{q,\ell}$ are real numbers.
Recent studies on the $\tau$ lepton and lepton electric/magnetic dipole moments provide bounds on the various leptonic couplings
\cite{Moyotl1,Moyotl2}.
For example, for $\Lambda_\calU=1$ TeV and $\dU=1.9$ the relevant couplings can be large as $\gtrsim\calO(1)$.
In this analysis we concentrate on the range $0\le c_i\le 1$.
As will be clear later, the unparticle contributes in the form of $(c_q\cdot c_\ell) (m_{B_s}^2/\Lambda_\calU^2)^\dU$,
thus larger values of $c_{q,\ell}$ could be compensated by larger $\dU$.
\par
We do not consider vector unparticle contributions because they are expected to be highly suppressed.
One can infer from Eq.\ (\ref{LU}) that the scalar unparticle contribution is proportional to 
$\sim (1/\Lambda_\calU^2)^{\dU}$, or more exactly (as will be shown later), $\sim(m_{B_s}^2/\Lambda_\calU^2)^\dU$.
On the other hand, the vector unparticle $\calO_\calU^\mu$ couples to the SM current as
\begin{equation}
 \frac{c_V}{\Lambda_\calU^{d_V-1}}\bar q'\gamma_\mu(1-\gamma_5)q~\calO_\calU^\mu~,
\end{equation}
where $d_V$ is the scaling dimension of $\calO_\calU^\mu$, and its contribution is $\sim (m_{B_s}^2/\Lambda_\calU^2)^{d_V-1}$.
But the unitarity constraints require that $\dU\ge 1$ and $d_V\ge 3$ \cite{unitarity},
resulting in much more suppression of the vector contribution \cite{jplee}.
\par
The total decay rate of $B_s\to\mu^+\mu^-$ is now given by
\begin{equation}
 \Gamma
=\frac{1}{16\pi M_{B_s}}\sqrt{1-\frac{4m_\mu^2}{m_{B_s}^2}}~
\Big|\langle\mu\mu|(\calH_{eff}^{\SM}+\calH_{eff}^\calU)|B_s\rangle\Big|^2~,
\end{equation}
where $\calH_{eff}^{\SM}$ is the SM effective Hamiltonian, while the unparticle effective Hamiltonian $\calH_{eff}^\calU$ is
\begin{equation}
 \calH_{eff}^\calU
=\frac{A_\dU e^{-i\phi_\calU}}{\sin\dU\pi}\left(\frac{m_{B_s}}{\Lambda_\calU}\right)^{2\dU}
   \left(\frac{m_\mu m_b}{m_{B_s}^4}\right)(c_q\cdot c_\ell)
   \big[{\bar b}(1-\gamma_5)s\big]\big[{\bar\ell}\gamma_5\ell\big]~.
\label{HU}
\end{equation}
Now the branching ratio can be written as
\begin{equation}
 \Br(B_s\to\mu\mu)=\Br_{\SM}\cdot|P|^2~.
\label{Br}
\end{equation}
Here $\Br_\SM$ is the SM prediction and
\begin{equation}
 P=1+\frac{m_{B_s}^2}{2m_\mu}\frac{m_b}{m_b+m_s}\frac{C_P}{C_{10}^{\rm SM}}~.
\label{P}
\end{equation}
The coefficients $C_{10}^{\rm SM}$ and $C_P$ are given by
\begin{eqnarray}
C_{10}^{\rm SM}&=&
  -\frac{1}{\sin^2\theta_W}\eta_Y Y_0(x_t)~,\label{C10}\\
C_P&=&
  \frac{\sqrt{2}\pi}{G_F\alpha(V_{tb}V_{ts}^*)}\frac{A_\dU e^{i\phi_\calU}}{\sin\dU\pi}
  \left(\frac{m_{B_s}}{\Lambda_\calU}\right)^{2\dU}
  \left(\frac{2m_\mu}{m_{B_s}^4}\right)(c_q\cdot c_\ell)^*~,\label{CP}
\end{eqnarray}
where $x_t=m_t^2/m_W^2$, $Y(x)=\eta_Y Y_0(x)$, and
\begin{equation}
 Y_0(x)=\frac{x}{8}\left[\frac{x-4}{x-1}+\frac{3x}{(x-1)^2}\ln x\right]~,~~~
\eta_Y=1.0113~.
\end{equation}
Some remarks are in order.
Our effective Lagrangian $\calL_\calU$ in Eq.\ (\ref{LU}) contains minimal couplings,
so the effective Hamiltonian $\calH_{eff}^\calU$ in Eq.\ (\ref{HU}) is just proportional to the leptonic pseudoscalar operator 
with left-handed quark sector.
If we added right-handed quark current in $\calL_\calU$, we would have leptonic pseudoscalar operator with right-handed quarks.
The corresponding coefficient is usually $C'_P$ in the literature, which is a counterpart of $C_P$ in Eq.\ (\ref{CP}).
As shown in \cite{Becirevic}, $C_P$ and $C'_P$ appear with different combinations in $B_s\to\ell^+\ell^-$ and $B\to K\ell^+\ell^-$ decays.
In $B_s\to\mu^+\mu^-$, the new physics contributes with $C_P-C'_P$ while in $B\to K\ell^+\ell^-$ with $C_P+C'_P$,
thus the two decay modes are complementary.
Numerically, one can estimate from Eqs.\ (\ref{EXP}), (\ref{Br}) and (\ref{P}) that 
(neglecting the nonzero $\Delta\Gamma_s$ effects discussed later)
\begin{equation}
|(C_P-C_P')m_b-0.16|=0.15~,
\label{C_P}
\end{equation}
while $|(C_P+C_P')m_b-0.33|\le 1.3$ from $B\to K\ell^+\ell^-$ \cite{Becirevic}.
Thus the new measurement of $B_s\to\mu^+\mu^-$ is very impressive for pinning down the Wilson coefficients.
\par
On the other hand in $B\to K^*\ell^+\ell^-$, new physics enters in $C_P-C'_P$ combination and 
the pseudoscalar operators are numerically irrelevant \cite{Altmannshofer}.
Estimation of Eq.\ (\ref{C_P}) is much smaller than the values considered in \cite{Altmannshofer}, 
$-0.38\lesssim (C_P-C_P')m_b\lesssim 0.63$,
so we expect that numerically $C_P^{(')}$ would be much more irrelevant to $B\to K^*\ell^+\ell^-$.
In the inclusive decay $B\to X_s\mu^+\mu^-$, the coefficients contribute as $|C_P|^2+|C'_P|^2$, 
which can be complementary to $B_s\to\mu^+\mu^-$ decay.
The constraint is rather weak however, since $m_b^2(|C_P|^2+|C_P'|^2)<45$ from $B\to X_s\mu^+\mu^-$ \cite{Alok}.
\par
In Table \ref{T1}, we summarize the input values used in this analysis.
\begin{table}
\begin{tabular}{ll}\hline
$G_F=1.16638\times 10^{-5}~{\rm GeV}^{-2}$ &~~~~~~~~~~~~~~~$\sin^2\theta_W=0.23116$ \\
$\alpha^{-1}=127.937$ &~~~~~~~~~~~~~~~ $V_{tb}=0.999$\\
$|V_{ts}|=0.0407$ &~~~~~~~~~~~~~~~$\phi_{ts}=-3.123$\\
$m_\mu=105.658$ MeV &~~~~~~~~~~~~~~~$m_t=163.2$ GeV\\
$m_{B_s}=5.3667$ GeV &~~~~~~~~~~~~~~~ $\tau_{B_s}=1.497$ ps\\
$f_{B_s}=234$ MeV &~~~~~~~~~~~~~~~$\Lambda_\calU=1000$ GeV\\
\hline
\end{tabular}
\caption{Input parameters used in this paper.
Here $\alpha^{-1}=\alpha(m_Z)^{-1}$, $m_t=m_t(m_t)$ in the ${\overline{\rm MS}}$ scheme, and
$\phi_{ts}$ is the phase of $V_{ts}$.}
\label{T1}
\end{table}
With the values of Table \ref{T1}, one gets $\Br_\SM=3.54\times 10^{-9}$, which is consistent with other literatures.
To compare the theoretical prediction with the experimental result, one should consider the non-zero decay width effect of $B_s$ meson
\cite{Genon,DeBruyn1,DeBruyn2}.
According to \cite{DeBruyn1}, 
\begin{equation}
 \Br(B_s\to\mu^+\mu^-)_{\rm theo}
=\left[\frac{1-y_s^2}{1+y_s\calA_{\Delta\Gamma}}\right]
   \Br(B_s\to\mu^+\mu^-)_{\rm exp}~,
\label{ysmod}
\end{equation}
where
\begin{equation}
 y_s\equiv \tau_{B_s}\frac{\Delta\Gamma_s}{2}=0.088\pm0.014~.
\end{equation}
Here
\begin{equation}
 \calA_{\Delta\Gamma}\equiv\frac{R_H-R_L}{R_H+R_L}~,
\end{equation}
where $R_{H(L)}\exp\left[-\Gamma^{(s)}_{H(L)}t\right]$ is the decay rate of the heavy (light) mass eigenstate.
In our case of Eq.\ (\ref{HU}) (pseudoscalar leptonic operator),
one can easily find that
\begin{equation}
 \calA_{\Delta\Gamma}=\cos(2\phi_P-\phi_s^{\rm NP})~,
\label{AGam}
\end{equation}
where $\phi_P$ is the phase of $P$ in Eq.\ (\ref{P}), 
and $\phi_s^{\rm NP}$ is the phase of new physics (in this case unparticles) in $B_s$-${\bar B}_s$ mixing.
From the analysis of \cite{jplee}, 
$\Delta=|\Delta|\cdot\exp(i\phi_s^{\rm NP})$ and
\begin{equation}
 \Delta=
1+\frac{1}{M_{12}^{\rm SM}}\frac{A_\dU e^{-i\phi_\calU}}{8\sin\dU\pi}
\left(\frac{f_{B_s}^2m_b^2}{m_{B_s}^3}\right)
\left(\frac{m_{B_s}}{\Lambda_\calU}\right)^{2\dU}
\frac{5c_q^2}{3}~,
\label{Delta}
\end{equation}
where $M_{12}^{\rm SM}$ is the standard model contribution.
\par
And the time-dependent CP asymmetric observable $\calS_{\mu\mu}$ is \cite{DeBruyn1,Zwicky}
\begin{equation}
 \calS_{\mu\mu}=\sin(2\phi_P-\phi_s^{\rm NP})~,
\label{Smumu}
\end{equation}
which is proportional to the helicity-summed time-dependent rate asymmetry,
$\calS_{\mu\mu}\sim \Gamma(B_s(t)\to\mu^+\mu^-)-\Gamma({\bar B}_s(t)\to\mu^+\mu^-)$.
\par
\begin{figure}
\begin{tabular}{cc}
\includegraphics[width=8cm]{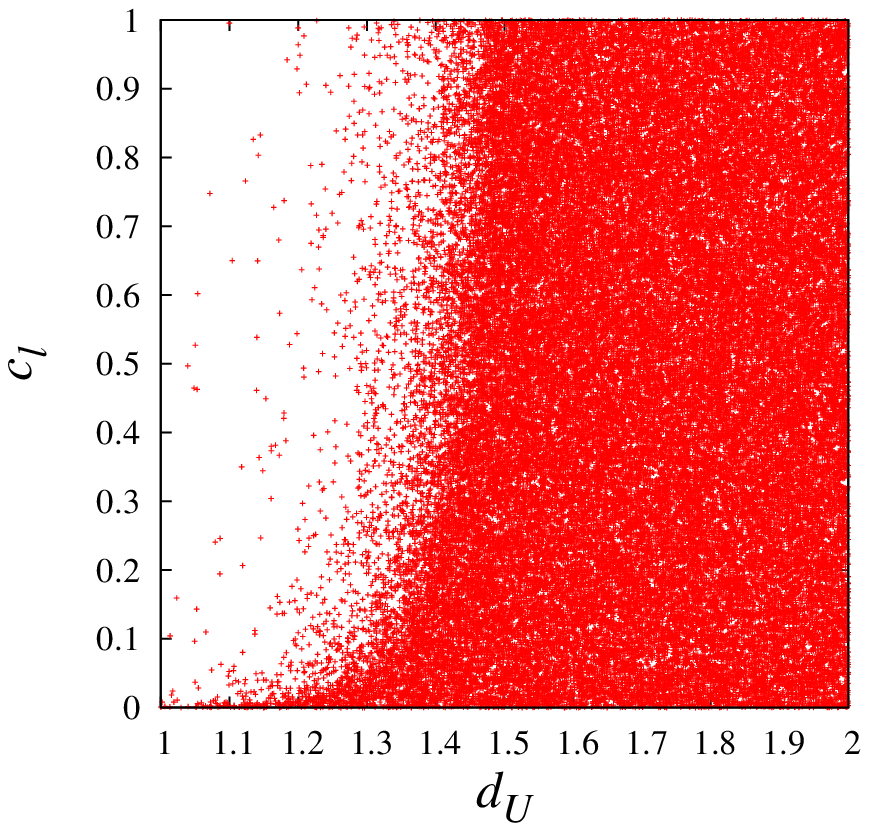}&
\includegraphics[width=8cm]{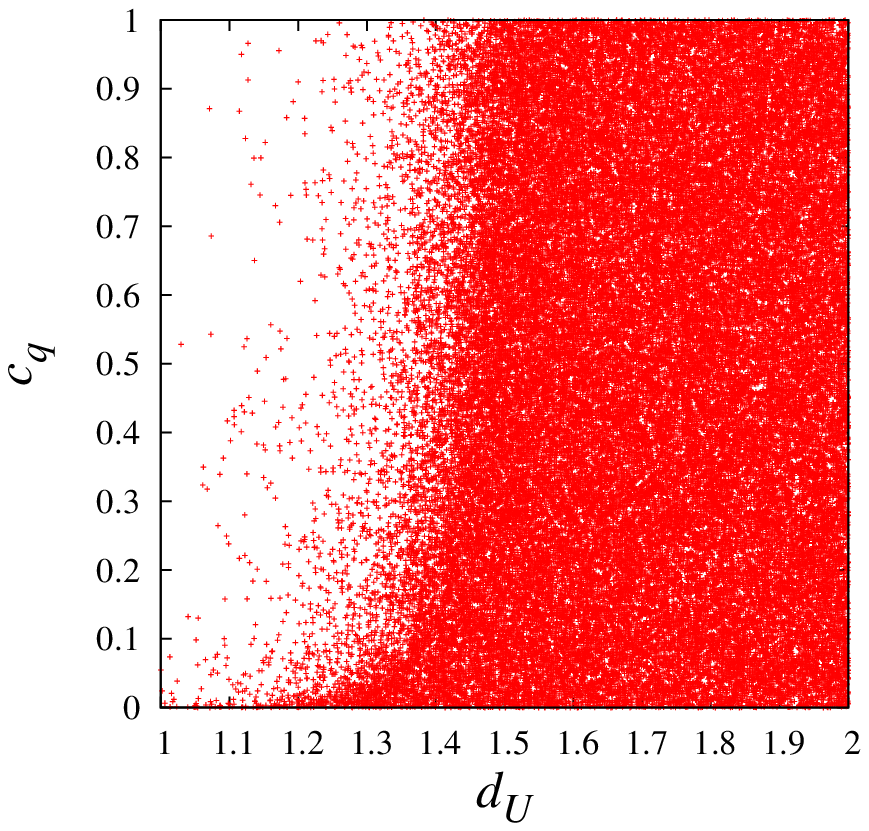}\\
(a) & (b)
\end{tabular}
\caption{Allowed region in $\dU$-$c_\ell$ and $\dU$-$c_q$ plane.}
\label{Fig1}
\end{figure}
\begin{figure}
\includegraphics{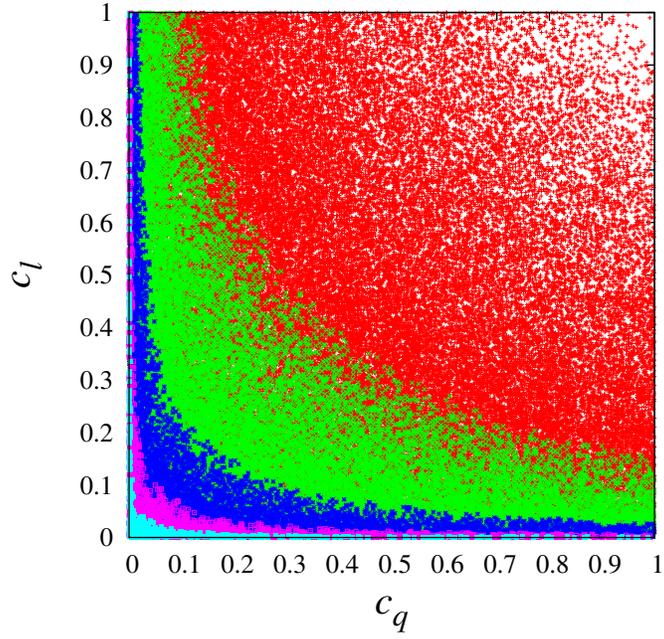}
\caption{Allowed values of $c_q$ and $c_\ell$ for $1.0\le\dU<1.1$ (cyan), $1.1\le\dU<1.2$ (pink), 
$1.2\le\dU<1.3$ (blue), $1.3\le\dU<1.4$ (green), and $1.4\le\dU<1.5$ (red).
For $\dU\gtrsim 1.5$, almost all the values of $0\le c_{q,\ell}\le 1.0$ are allowed.}
\label{Fig2}
\end{figure}

Figure \ref{Fig1} shows the allowed values of $c_q$ and $c_\ell$ versus $\dU$ constrained by the measured branching ratio, Eq.\ (\ref{EXP}).
The behavior of Fig.\ \ref{Fig1} can be inferred from the Eqs.\ (\ref{Br}) and (\ref{CP}).
Note that $C_P$ is proportional to $(m_{B_s}/\Lambda_\calU)^{2\dU}\simeq (2.88\times 10^{-5})^{\dU}$,
which suppresses the unparticle contribution to the total branching ratio significantly for $1<\dU<2$.
Thus for larger values of $\dU$, the value of $c_q\cdot c_\ell$ can be large to fit the experimental result.
For some combinations of $c_q$ and $c_\ell$, $\calA_{\Delta\Gamma}$ can be negative in Eq.\ (\ref{AGam}),
allowing rather smaller values of $\dU$.
As shown in Fig.\ \ref{Fig1}, for $\dU\gtrsim 1.4$, almost all the region of $0\le c_q\le 1$ or $0\le c_\ell\le 1$ is allowed.
Figure \ref{Fig2} shows the allowed region of $(c_q,c_\ell)$ for different values of $\dU\le 1.5$.
Note that the red points corresponding to $1.4\le\dU\le 1.5$ cover almost all the space of $0\le c_{q,\ell}\le 1.0$.
That is the reason why we do not consider the region $\dU>2$ in this analysis.
It was pointed out in \cite{Sannino} that the best candidate for the scalar operator is the fermion bilinear $\calO={\bar\psi}\psi$,  
and current lattice simulation indicates that the scaling dimension of this operator is larger than 2.
If that was the case, the scalar unparticle contribution gets very suppressed and the vector unparticle contributions might be 
comparable to the scalar ones.
In this analysis we are considering general scalar operator with scaling dimension $\dU\ge 1$.
\par
\begin{figure}
\includegraphics{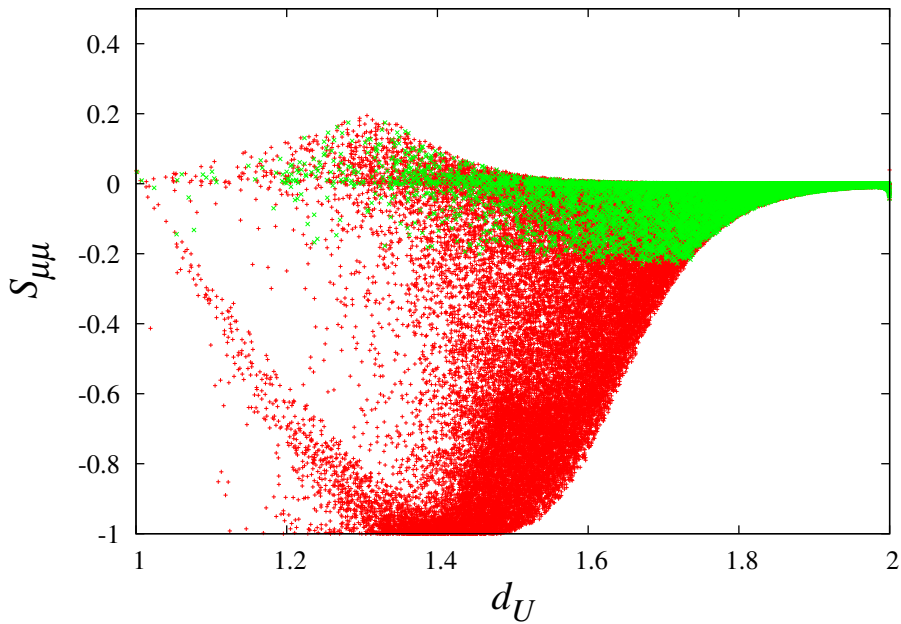}
\caption{Time-dependent CP asymmetry parameter $\calS_{\mu\mu}$ versus $\dU$.
Green points are from the constraints of $B_s\to\psi\phi$ while red ones are unconstrained.}
\label{Fig3}
\end{figure}

In Fig.\ \ref{Fig3} we show the time-dependent CP asymmetry parameter $\calS_{\mu\mu}$ as a function of $\dU$.
The figure shows that unconstrained $\calS_{\mu\mu}$ (red points) is mostly negative.
But if we impose the constraints from $B_s\to\psi\phi$ 
where $-0.20\le \calS_{\psi\phi}\equiv\sin(2|\beta_s|-\phi_s^{\rm NP})\le 0.20$ ($\beta_s$ is the phase of $V_{ts}$)
\cite{Buras:S}, then
$-0.25\lesssim\calS_{\mu\mu}\lesssim 0.2$ (green points).
Note that the $S_{\psi\phi}$ constraint is very strong.
If $\calS_{\mu\mu}$ turned out to be $|\calS_{\mu\mu}|\gtrsim 0.25$, it could not be explained by unparticles.
Figure \ref{Fig3} can be used to distinguish scalar unparticles from ordinary scalar particles.
It was shown in \cite{Buras:S} that $|\calS_{\mu\mu}|\lesssim 0.5$, 
for $C_P$ contributions from new scalar particles.
In \cite{Buras:S} non-zero phase of $\calS_{\mu\mu}$ comes from the complex couplings, 
but in this work the source of the phase is $\phi_\dU$ with real couplings.
If $\dU\to 1$ and the unparticle couplings are complex 
then contributions of the scalar unparticle become those of ordinary scalar particles,
since in this limit $C_P\to\sim c_q c_\ell/\Lambda_\calU^2$,
which is equivalent to the $C_P$ of \cite{Buras:S}.
\par
It should be noticed that only a replacement of 
$(m_{B_s}/\Lambda_\calU)^{2\dU}(1/\Lambda_\calU^2)$ with
$1/M_0^2$ where $M_0$ is a mass of some new scalar particle
is not enough to reduce the unparticle to an ordinary scalar particle,
because there is a nontrivial phase associated with $\dU$.

For non-integral $\dU$, it serves as a phase of new physics but it also suppresses new physics effects through
$(m_{B_s}/\Lambda_\calU)^{2\dU}$.
That's the reason why the allowed region of $\calS_{\mu\mu}$ from unparticles is smaller than that from ordinary particles.
This is a very unique feature of unparticles.
For ordinary particles, to suppress the new physics contributions the new couplings should be small or
the mass of new particle must be large.
But in the unparticle scenario, non-integral scaling dimension $\dU$ can do the work,
and $\dU$ itself enters as a new phase as shown in Eq.\ (\ref{propa}).
\par
Current analysis is done for $\Lambda_\calU=1$ TeV.
For larger values of $\Lambda_\calU$, unparticle contribution gets smaller by $(m_{B_s}/\Lambda_\calU)^{2\dU}$
and the allowed parameter space would become larger.
\par
In conclusion, we have investigated the unparticle effects on $B_s\to\mu^+\mu^-$ decay.
The experimental result is quite consistent with the SM, but it does not mean that there is no room for new physics.
In this analysis only the scalar unparticles are considered because vector unparticles are expected to give negligible contributions.
Assuming that scalar unparticles couple to the left-handed current, 
we provided the allowed regions of the couplings $c_{q,\ell}$ and the scaling dimension $\dU$ for a fixed $\Lambda_\calU=1~{\rm TeV}$.
Since the unparticle contributions are proportional to $(m_{B_s}/\Lambda_\calU)^{2\dU}$, 
the allowed parameter space of $c_{q,\ell}$ gets larger for large $\dU$.
The upper bound on $\Br(B_d\to\mu^+\mu^-)$ of Eq.\ (\ref{BdEXP}) would not give strong constraints on the model parameters,
since the SM prediction of Eq.\ (\ref{SMBd}) is almost an order of magnitude smaller.
But if the branching ratio $\Br(B_d\to\mu^+\mu^-)$ is measured in near future, a combined analysis with $B_s\to\mu^+\mu^-$ would
give some hints on the flavor structure of unparticle interactions.
And the scalar unparticle predicts mostly negative $\calS_{\mu\mu}$, which could be used to distinguish unparticles from 
ordinary particles.
\begin{acknowledgments}
The author thanks Kang Young Lee for his reminding of the subject and helpful discussions.
This work is supported by WCU program through the KOSEF funded by the MEST (R31-2008-000-10057-0).
\end{acknowledgments}

\end{document}